\def\rfr#1{eq. (\ref{#1})}

\def\derp#1#2{\rp{\partial{#1}}{\partial{#2}}}
\def\dert#1#2{\frac{{{d}}{#1}}{{{d}}{#2}}}
\def\virg#1{``#1''}

\def\eqi{\begin{equation}}
\def\eqf{\end{equation}}
\def\eqia{\begin{eqnarray}}
\def\eqfa{\end{eqnarray}}
\def\Om{\mathit{\Omega}}
\def\rp#1#2{{#1\over#2}} \def\lb#1{\label{#1}}

\def\bds#1{\boldsymbol{#1}}
\def\co{\cos\omega}
\def\so{\sin\omega}

\def\sI{\sin I}

\def\ton#1{\left(#1\right)}
\def\qua#1{\left[#1\right]}
\def\grf#1{\left\{#1\right\}}

\documentclass[useAMS,usenatbib]{mn2e}
\usepackage{amsmath}
\usepackage{amscd}
\usepackage{amssymb}
\usepackage{txfonts}
\RequirePackage{color}

\title[Local cosmological effects of the order of $H$ in orbits]{Local cosmological effects of the order of $H$ in the orbital motion of a binary system}

\author[L. Iorio]{L.
Iorio$^{1}$\thanks{E-mail:
lorenzo.iorio@libero.it}\\
$^{1}$I Ministero dell'Istruzione, dell'Universit\`{a} e della
Ricerca (M.I.U.R.), Viale Unit\`{a} di Italia 68
Bari, (BA) 70125,
Italy}

\begin{document}

\maketitle

\label{firstpage}

\begin{abstract}
A two-body system hypothetically affected by an additional radial acceleration $H v_r$, where $v_r$ is the radial velocity of the binary's proper orbital motion, would experience long-term temporal changes of both its semimajor axis $a$ and the eccentricity $e$ qualitatively different from any other standard competing effect for them. Contrary to what one might reasonably expect, the analytical expressions of such rates do not vanish in the limit $M\rightarrow 0,$ where $M$ is the mass of the primary, being independent of it. This is a general requirement that any potentially viable physical mechanism able to provide such a putative acceleration should meet. Nonetheless, if $H$ had the same value $H_0$ of the Hubble parameter at present epoch, such rates of change would have magnitude close to the present-day level of accuracy in determining planetary orbital motions in our Solar System.
%A tension with recent observations may even be present for Mercury and Mars.
However, general relativity, applied to a localized gravitationally bound binary system immersed in an expanding Friedmann-Lema\^{\i}tre-Robertson-Walker, does not predict the existence of such a putative radial acceleration at Newtonian level. Instead, it was recently shown in literature that an acceleration of order $H$ and directed along the velocity $\bds v$ of the test particle occurs at post-Newtonian level. We worked out its orbital effects finding well-behaved secular rates of change for both $a$ and $e$ proportional to the Schwarzschild radius $r_s$ of the primary. Their magnitude is quite small: the rate of change of $a$ amounts to just 20 $\mu$m per century in our Solar System. Finally, we discussed certain basic criteria of viability that modified models of gravity should generally  meet when their observable effects are calculated.
\end{abstract}

%\keywords{gravitation-Oort Cloud-celestial mechanics}
%\keywords{black hole physics-Galaxy:center-relativity-techniques: radial velocities}
%\keywords{Experimental studies of gravity; Experimental tests of gravitational theories; Ephemerides, almanacs, and calendars}
%PACS: 04.80.-y; 04.80.Cc; 95.10.Km
%
%\keywords{gravitation - celestial mechanics - ephemerides - planet$-$disc interactions - Kuiper belt: general - minor planets, asteroids: general}
%\centerline
%{PACS: 04.80.-y; 04.80.Cc; 98.80.Es; 95.10.Km}
%


 \maketitle

\section{Introduction}
In this paper, we first deal with a  certain hypothetical anomalous radial acceleration  proportional to the radial velocity of the orbital motion of a two-body system through a coefficient $H$ having dimensions of ${\rm T}^{-1}$.

Some intuitive, Newtonian-like guesses about the possibility that such a putative acceleration may exist as a local manifestation of the cosmic expansion in the case of non-circular motions are offered in Section \ref{conget}. They are motivated by the known fact that, within certain limits, several well established key features of a homogeneous and isotropic expanding universe, and also of its influence on local gravitationally bound systems, can be practically inferred within a classical framework \citep{1965AnPhy..35..437H,1996MNRAS.282..206T,2010RvMP...82..169C,2012arXiv1203.5596B,2012arXiv1207.0060F}.

Such a putative extra-acceleration may, in principle,  have interesting phenomenological consequences  since it would induce peculiar orbital signatures which would not be mimicked by any other known competing dynamical effect. Moreover, by assuming for $H$ a value equal to that of the Hubble parameter at present epoch, the magnitude of these exotic effects for the planets of our Solar System would be close to the current level of accuracy in determining their orbits.
Tensions might even occur between data and predictions in the case of Mercury and Mars. These topics are treated in Section \ref{osser}

In Section \ref{vaffa} we try-unsuccessfully-to find a theoretical justification for the guesses of Section \ref{conget} rooted in a full general relativistic treatment of the orbital dynamics of a local system embedded in an expanding homogeneous and isotropic Friedmann-Lema\^{\i}tre-Robertson-Walker (FLRW) spacetime metric.  Nonetheless, our results remain valid from a phenomenological point of view because of their actual independence of any specific theoretical scheme, and can be viewed as observational constraints on such a putative exotic force, whatever the physical mechanism yielding it (if any) may be. Moreover, we feel that the numerical values coming out from our analysis  are interesting if compared with the observations, and may pursue further investigations to find a possible physical origin, cosmological or not, for it. Cosmological effects linear in $H$ were recently derived by Kopeikin \citep{2012arXiv1207.3873K} for the propagation of electromagnetic waves between atomic clocks in geodesic motion in a FLRW background.

In Section \ref{conside}  we build on the certain aspects discussed in previous Sections and provide some very general viability criteria that must be met by modified models of gravity \citep{2012PhR...513....1C} in order not to give rise to unphysical observable effects. In particular, it is important to check the behaviour of their detectable predictions in the limits $G\rightarrow 0,M\rightarrow 0,$ where $G$ is the Newtonian constant of gravitation and $M$ is the mass of the central body acting as localized source of the gravitational field.

Section \ref{conclusioni} provides an overview of the results obtained.
\section{A local cosmological effect linear in $H$ for two-body orbital dynamics?}\lb{conget}
Let us define the Hubble parameter $H$ and the deceleration parameter $q$ in the usual way as
\begin{align}
H \lb{ho}& \doteq \rp{\dot S}{S}, \\ \nonumber \\
q \lb{qo}&\doteq -\rp{1}{H^2}\ton{\rp{\ddot{S}}{S}},
\end{align}
where $S(t)$ is the cosmological scale factor. From an observational point of view, the value of the Hubble parameter at present epoch is \citep{2011ApJ...730..119R} \eqi H_0 \doteq 100\ {\rm km\ s^{-1}\ Mpc^{-1}} h = (73.8\pm 2.4)\ {\rm km\ s^{-1}\ Mpc^{-1}},\eqf
so that
\eqi h = 0.738 \pm 0.024\lb{HU}.\eqf
The deceleration parameter can be connected to directly determined quantities from observations by means of \citep{2010obco.book.....S}
\eqi q_0=\rp{\Omega_{\rm m,0}}{2}-\Omega_{\Lambda,0},\eqf where $\Omega_{\rm m,0}$ and $\Omega_{\Lambda,0}$ are the current matter and dark energy densities, respectively, in units of the critical density \citep{2010obco.book.....S}
\eqi\rho_0^{\rm crit}=\rp{3H_0^2}{8\pi G}.\lb{critd}\eqf
From \citep{2011ApJS..192...14J}
\begin{align}
\Omega_{\rm m,0}h^2 &= 0.1334^{+0.0056}_{-0.0055}, \\ \nonumber \\
\Omega_{\Lambda,0} &= 0.728^{+0.015}_{-0.016},
\end{align}
and \rfr{HU}, it turns out
\eqi q_0\sim -0.7.\eqf

Let us now consider a two-body system composed by a central object of mass $M$ and a test particle gravitationally bound to it. Their proper motion is determined by their mutual gravitational interaction according to the Newtonian/Einsteinian laws. It superimposes on the global Hubble flow in such a way that the radial velocity of the test particle is the sum of such two contributions
\eqi v_r \doteq \dot r = v_r^{\rm (orb)} + H r.\eqf In it, $r$ is their relative distance and
\eqi v_r^{\rm (orb)}=\rp{n_{\rm b}ae\sin f}{\sqrt{1-e^2}}\lb{vkep}\eqf is the radial component of the velocity vector $\bds v$ of the test particle along its standard two-body Keplerian ellipse  where
$a$ is the semimajor axis, $e$ is the eccentricity, $f$ is the true anomaly, and $n_{\rm b}\doteq \sqrt{GM/a^3}$ is the Keplerian mean motion.
As shown by Table \ref{tavola0}, the contribution of the Hubble flow to the planetary radial velocities is quite negligible in the Solar System.
\begin{table*}
\caption{Keplerian two-body and cosmological radial velocities of the planets of the Solar System, in km s$^{-1}$. As far as the Keplerian two-body velocities are concerned, the maximum values of \rfr{vkep} ($\sin f=1$) were taken. The average planetary distances $d=a\ton{1+\rp{e^2}{2}}$ were used in the Hubble terms which were evaluated at the present epoch according to \citep{2011ApJ...730..119R} $H_0=73.8\ {\rm km\ s^{-1}\ Mpc^{-1}}= 2.39\times 10^{-18}\ {\rm s}^{-1}$.
}\label{tavola0}
\centering
\bigskip
\begin{tabular}{lll}
\hline\noalign{\smallskip}
Planet & $v_r^{\rm(orb)}$ $\ton{{\rm km\ s^{-1}}}$ & $ H_0 d$ $\ton{{\rm km\ s^{-1}}}$  \\
\noalign{\smallskip}\hline\noalign{\smallskip}
Mercury & $10.05$ & $1\times 10^{-10}$ \\
Venus & $0.23$ & $2\times 10^{-10}$ \\
Earth & $0.49$ & $3\times 10^{-10}$ \\
Mars & $2.26$ & $5\times 10^{-10}$ \\
Jupiter & $0.59$ & $2\times 10^{-9}$ \\
Saturn & $0.52$ & $3\times 10^{-9}$ \\
Uranus & $0.31$ & $7\times 10^{-9}$ \\
Neptune & $0.04$ & $1\times 10^{-8}$\\
Pluto & $1.22$ & $1\times 10^{-8}$ \\
\noalign{\smallskip}\hline\noalign{\smallskip}
\end{tabular}
\end{table*}

When the accelerations are computed, neglecting the proper motion yields the well known Hooke-like term quadratic in $H$. Indeed, starting from the Hubble law
\eqi \dot r = H r,\lb{appro}\eqf
yields
\eqi A_r = \dot H r + H\dot r.\lb{accela}\eqf
By recalling that
\eqi\dot H=\rp{\ddot{S}}{S} -H^2\eqf and by using \rfr{qo} and \rfr{appro}
one gets just
\eqi A_r^{\ton{H^2\rm N}} = -q H^2 r.\lb{accel2}\eqf
If we include $v_r^{(\rm orb)}$ in the second term of \rfr{accela}, we get an additional term linear in $H$
\eqi A_r^{(H\rm N)}= H v_r^{(\rm orb)}\lb{accel}\eqf at Newtonian level.

\section{A cosmological test particle acceleration of order $H v_r$: a phenomenologically appealing possibility}\lb{osser}
Table \ref{tavola00} shows the values of $A_r^{(H\rm N)}$ and $A_r^{\ton{H^2\rm N}}$ compared with the Newtonian monopoles for the planets of the Solar System.
\begin{table*}
\caption{Newtonian and cosmological accelerations of the planets of the Solar System, in m s$^{-2}$. As far as the Hubble  terms of order $H$ are concerned, the maximum values of \rfr{vkep} ($\sin f=1$) were taken. The average planetary distances $d=a\ton{1+\rp{e^2}{2}}$ were used in both the Newtonian and the Hubble terms of order $H^2$ which were evaluated at the present epoch according to \citep{2011ApJ...730..119R} $H_0=73.8\ {\rm km\ s^{-1}\ Mpc^{-1}}= 2.39\times 10^{-18}\ {\rm s}^{-1}$ and \citep{2011ApJ...730..119R} $q_0 \sim -0.7$.
}\label{tavola00}
\centering
\bigskip
\begin{tabular}{llll}
\hline\noalign{\smallskip}
Planet & $GM/d^2$ $\ton{{\rm m\ s^{-2}}}$ & $H_0 v_r^{\rm(orb)}$ $\ton{{\rm m\ s^{-2}}}$ & $ -q_0 H_0^2 d$ $\ton{{\rm m\ s^{-2}}}$   \\
\noalign{\smallskip}\hline\noalign{\smallskip}
Mercury & $4\times 10^{-2}$ & $2.40\times 10^{-14}$ & $2\times 10^{-25}$ \\
Venus & $1\times 10^{-2}$ & $5\times 10^{-16}$ & $4\times 10^{-25}$\\
Earth & $6\times 10^{-3}$ & $1.2\times 10^{-15}$ & $5\times 10^{-25}$\\
Mars & $2\times 10^{-3}$ & $5.4\times 10^{-15}$ & $8\times 10^{-25}$\\
Jupiter & $2\times 10^{-4}$ & $1.4\times 10^{-15}$ & $2.6\times 10^{-24}$\\
Saturn & $6\times 10^{-5}$ & $1.2\times 10^{-15}$ & $4.9\times 10^{-24}$\\
Uranus & $2\times 10^{-5}$ & $7\times 10^{-16}$ & $9.8\times 10^{-24}$\\
Neptune & $6\times 10^{-6}$ & $1\times 10^{-16}$ & $1.54\times 10^{-23}$\\
Pluto & $4\times 10^{-6}$ & $2.9\times 10^{-15}$ & $2.09\times 10^{-23}$ \\
\noalign{\smallskip}\hline\noalign{\smallskip}
\end{tabular}
\end{table*}
It can be noticed that the orders of magnitude of $A_r^{(H\rm N)}$ and $A_r^{\ton{H^2\rm N}}$ are completely different, being the terms linear in $H$ of the order of $10^{-14}-10^{-16}$ m s$^{-2}$. Interestingly, such values are neither too large to unrealistically compromise the agreement between theory and observations nor too small to be completely undetectable in any foreseeable future. Indeed, according to a recent analysis of various kind of planetary data by Folkner \citep{folkner09}, the largest unmodelled radial acceleration in the Solar System allowed by observations is just of the order of $10^{-14}$ m s$^{-2}$. More precisely, from the current Mars radio range data set, Folkner \citep{folkner09} found upper bounds on a radial acceleration of Earth and Mars to be
less than $3\times 10^{-14}$ m s$^{-2}$ and $8\times 10^{-14}$ m s$^{-2}$, respectively. From Cassini radiotechnical data Folkner \citep{folkner09} inferred an upper bound of $1\times 10^{-14}$ m s$^{-2}$ for Saturn. Thus, according to Table \ref{tavola00}, $A_r^{(H\rm N)}$ seems to be not too far from the edge of the present-day detectability. Moreover, the maximum values for $A_r^{(H\rm N)}$ quoted in Table \ref{tavola00} are not in contrast with the upper bounds by Folkner \citep{folkner09} for Earth, Mars and Saturn. On the other hand, it must be remarked that Folkner \citep{folkner09} did not release details about the time and/or spatial variability of the anomalous acceleration constrained. This fact generally does matter since $A_r^{(H\rm N)}$ is not constant, so that the constraints by Folkner \citep{folkner09} may not be straightforwardly applicable to $A_r^{(H\rm N)}$.

We now explicitly work out some dynamical orbital effects induced by \rfr{accel} on some quantities routinely determined from observations by astronomers.
The standard Gauss equations \citep{befa} for the variations of the Keplerian orbital elements, applied to \rfr{accel} evaluated onto the unperturbed Keplerian ellipse by means of \rfr{vkep}, yield the following non-zero long-term rates of change of the semimajor axis $a$ and the eccentricity $e$
\begin{align}
\dot a^{(H\rm N)}\lb{ratea} &= 2a H_0 \ton{1-\sqrt{1-e^2}}= a  H_0 e^2 + \mathcal{O}\ton{e^4}, \\ \nonumber \\
\dot e^{(H\rm N)}\lb{ratee} &=  \rp{H_0\ton{1-e^2}\ton{1-\sqrt{1-e^2}}}{e} = \rp{H_0 e}{2}\ton{1 - \rp{3}{4}e^2} +\mathcal{O}\ton{e^4}.
\end{align}
It is intended that \rfr{ratea}-\rfr{ratee} are averages over one full orbital revolution of the test particle.
The other orbital elements are left unaffected.
Note that, according to \rfr{ratea}-\rfr{ratee}, both the  semimajor axis and the eccentricity increase.
This implies that the  mean distance $d = a\ton{1+{e^2}/{2}}$ increases at a rate
\eqi \dot d^{(H\rm N)} = 3 a H_0\ton{1-\sqrt{1-e^2}}=\rp{3a H_0 e^2}{2} + \mathcal{O}\ton{e^2}.\lb{dst}\eqf
Rather surprisingly, \rfr{ratea}-\rfr{dst}, which describe small changes occurring in the otherwise close orbit of a gravitationally bound system, do not depend on $M$: in the limit $M\rightarrow 0$, \rfr{ratea}-\rfr{dst} do not vanish. The limit $G\rightarrow 0$ does not pose problems in the sense that \rfr{ratea}-\rfr{dst} correctly vanish. Indeed, in a spatially flat  FLRW universe, it is \citep{2010obco.book.....S}
\eqi H^2 = \rp{8\pi G\rho}{3},\lb{cosmo1}\eqf where $\rho$ is the density of the cosmic fluid, inclusive of
any dark energy component.
In Table \ref{tavola} we calculate \rfr{ratea}-\rfr{ratee} for the planets of the Solar System.
\begin{table*}
\caption{Predicted cosmological rates of change of the semimajor axis $a$ and the eccentricity $e$ of the planets of the Solar System for \citep{2011ApJ...730..119R} $ H_0 = (73.8\pm 2.4)\ {\rm km\ s^{-1}\ Mpc^{-1}}= (2.39\pm 0.07)\times 10^{-18}\ {\rm s}^{-1} = (7.5\pm 0.2)\times 10^{-9}\ {\rm cty}^{-1}$ in \rfr{ratea}-\rfr{ratee}. The rates of $a$ for Mercury and Mars, and the rate of $e$ for Mars are larger by one order of magnitude than the corresponding experimental uncertainties in Table \ref{tavolapit}.
}\label{tavola}
\centering
\bigskip
\begin{tabular}{lll}
\hline\noalign{\smallskip}
Planet & $\dot a^{(H\rm N)}$ $\ton{{\rm m}\ {\rm cty^{-1}}}$ & $\dot e^{(H\rm N)}$ $\ton{{\rm cty^{-1}}}$ \\
\noalign{\smallskip}\hline\noalign{\smallskip}
\textcolor{black}{Mercury} & $\textcolor{black}{18.68\pm 0.61}$ & $(7.5\pm 0.2)\times 10^{-10}$  \\
Venus & $0.03\pm 0.001$ & $(2\pm 0.08)\times 10^{-11}$ \\
Earth & $0.31\pm 0.01$ & $(6\pm 0.2)\times 10^{-11}$  \\
\textcolor{black}{Mars} & $\textcolor{black}{15.04\pm 0.49}$ & $\textcolor{black}{(3.5\pm 0.1)\times 10^{-10}}$  \\
Jupiter & $11.94\pm 0.38$ & $(1.7\pm 0.05)\times 10^{-10}$ \\
Saturn & $31.26\pm 1.01$ & $(2.0\pm 0.06)\times 10^{-10}$ \\
Uranus & $46.56\pm 1.51$ & $(1.7\pm 0.05)\times 10^{-10}$ \\
Neptune & $2.47\pm 0.08$ & $(3\pm 0.1)\times 10^{-11}$ \\
Pluto & $2849.75\pm 92.67$ & $(9.0\pm 0.3)\times 10^{-10}$ \\
\noalign{\smallskip}\hline\noalign{\smallskip}
\end{tabular}
\end{table*}
It is interesting to compare the values in Table \ref{tavola} with the experimental bounds in Table \ref{tavolapit}, preliminarily inferred from a multi-year fit by Pitjeva \citep{pitjeva07} for the EPM2006 ephemerides.
\begin{table*}
\caption{Uncertainties in the rates of change of the semimajor axis $a$ and the eccentricity $e$ of the planets of the Solar System. They were inferred  by taking the ratios of the formal errors in Table 3 of \citep{pitjeva07}, all rescaled by a factor 10, to the data time span $\Delta T=93$ yr (1913-2006) of the EPM2006 ephemerides used by Pitjeva \citep{pitjeva07}.  The results for Saturn are relatively inaccurate with respect to those of the inner planets since radiotechnical data from Cassini were not yet processed when Table 3 of \citep{pitjeva07} was produced. Here cty stands for century.
}\label{tavolapit}
\centering
\bigskip
\begin{tabular}{lll}
\hline\noalign{\smallskip}
Planet & $\sigma_{\dot a}$ $\ton{{\rm m}\ {\rm cty^{-1}}}$ & $\sigma_{\dot e}$ $\ton{{\rm cty^{-1}}}$  \\
\noalign{\smallskip}\hline\noalign{\smallskip}
\textcolor{black}{Mercury} & $\textcolor{black}{3.6}$ & $4\times 10^{-9}$ \\
Venus & $2.3$ & $2\times 10^{-10}$ \\
Earth & $1.5$ & $5\times 10^{-11}$ \\
\textcolor{black}{Mars} & $\textcolor{black}{2.8}$ & $\textcolor{black}{5\times 10^{-11}}$ \\
Jupiter & $6612.9$ & $2\times 10^{-8}$ \\
Saturn & $45763.4$ & $1\times 10^{-7}$ \\
Uranus & $433269.0$ & $3\times 10^{-7}$ \\
Neptune & $4.9818\times 10^6$ & $8\times 10^{-7}$\\
Pluto & $3.66961\times 10^7 $ & $3\times 10^{-6}$ \\
\noalign{\smallskip}\hline\noalign{\smallskip}
\end{tabular}
\end{table*}
In general, the predicted rates of Table \ref{tavola} are compatible with the bounds of Table \ref{tavolapit}, although discrepancies occur for Mercury and Mars at $4-\sigma$ level. It must be stressed that, so far, astronomers did not explicitly determine corrections to the standard Newtonian rates of change of $a$ and $e$ from observations: the figures in Table \ref{tavolapit} were inferred rather naively by simply taking the ratios of the formal, statistical uncertainties in the orbital elements,  rescaled by a factor 10, to the time span of the fit performed by Pitjeva \citep{pitjeva07}. In view of future
analyses, it is important to remark that \rfr{ratea}-\rfr{ratee}, assumed as real physical effects, show very distinctive patterns since there are no other known competing orbital effects on $a$ and $e$, both Newtonian and Einsteinian. Indeed, general relativity does not predict any long-term rates of change for such orbital elements: the Schwarzschild field of a spherical static body affects just the pericenter and the mean anomaly of a test particle, while the Lense-Thirring effect due to the rotation of the central body causes secular precessions of the node and the pericenter. As far as Newtonian mechanics is concerned, the centrifugal oblateness of the primary does not impact $a$ and $e$. Long-term rates of change of the eccentricity may be induced by unmodeled/mismodeled distant distributions of matter like a massive ring or a pointlike object having certain specific orbital configurations, but their temporal signatures are known, and are quite different with respect to \rfr{ratee} so that it would be relatively simple to separate them. An isotropic mass loss suffered by the central body actually causes long-term variations of $a$ and $e$, but, in the case of the Sun, they are smaller than the predicted cosmological values in Table \ref{tavola}. In any case, they have  different patterns with respect to \rfr{ratea}-\rfr{ratee}. Moving to non-standard effects, in principle, a violation of the strong equivalence principle and spacetime variations of fundamental constants\footnote{A time-dependent $G(t)$ would change $a$ as well, but not as predicted by \rfr{ratea}.} like, e.g., $G$ can impact the eccentricity, but, again, not in the same way as predicted by \rfr{ratee}.

In the following we will calculate the shifts per orbit $\Delta Y$ of various standard observables $Y$ as
\begin{align} \Delta Y \lb{DY} \nonumber &  =\int_0^{P_{\rm b}}dY = \int_0^{P_{\rm b}}\ton{\dert Y t}dt = \\ \nonumber \\
&= \int_0^{2\pi}\qua{\derp Y{\textcolor{black}{f}}\dert {\textcolor{black}{f}}{\mathcal{M}}\dert{\mathcal{M}}t + \sum_{\kappa}\derp Y\kappa\dert\kappa t}\ton{\dert t{\textcolor{black}{f}}}d\textcolor{black}{f},\ \kappa = a,e,I,\omega,\Om, \end{align}
where $P_{\rm b}=2\pi/n_{\rm b}$ is the orbital period along the  Keplerian ellipse, and $\mathcal{M},I,\omega,\Om$ are the mean anomaly, the inclination, the argument of pericenter and the longitude of the ascending node, respectively, of the orbit of the test particle. $Y$ appearing in \rfr{DY} is the analytical expression for the observable $Y$ evaluated onto the unperturbed Keplerian orbit. $d\mathcal{M}/dt$ and $d\kappa/dt, \kappa = a,e,I,\omega,\Om$ in \rfr{DY} are the instantaneous rates of change of the  Keplerian orbital elements given by the right-hand-sides of the Gauss equations evaluated onto the unperturbed Keplerian  ellipse. Expressions for $df/d\mathcal{M},dt/df$ can be found in standard textbooks.

The shift per orbit of the radial distance
\eqi r=\rp{a\ton{1-e^2}}{1 + e\cos f}\eqf turns out to be
\eqi \Delta r^{(H\rm N)} \lb{range}  = -\rp{3\pi a H_0\ton{1-e^2}\ton{-2 + e^2 +2\sqrt{1-e^2}}}{n_{\rm b} e^2} = -\rp{3\pi a H_0 e^2}{4 n_{\rm b}} + \mathcal{O}\ton{e^4}.\eqf

The radial velocity is left unaffected.

The shift per orbit of the line-of-sight projection of the orbit of a binary system in the sky
\eqi\rho = r\sI\sin\ton{\omega + f},\eqf which is practically determined from timing measurements of compact objects, is
\begin{align}\Delta\rho^{(H\rm N)} \lb{los}\nonumber & = -\rp{3\pi a H_0\sI\so\ton{1-e^2}\ton{-2 + e^2 +2\sqrt{1-e^2}}}{n_{\rm b}e^3} = \\ \nonumber \\
& =\rp{3\pi ae H_0\sI\so}{4n_{\rm b}}\ton{1 -\rp{e^2}{2} } + \mathcal{O}\ton{e^4}.\end{align}
where, in this case, $I$ is the inclination of the orbital plane to the plane of the sky.
Curiously, both \rfr{range} and \rfr{los} are independent of $G$ because of \rfr{cosmo1}, while they depend on $M$ in such a way that they diverge in the limit $M\rightarrow 0$. Such a singularity might be explained by noticing that both \rfr{range} and \rfr{los} are the outcome of a perturbative calculation in which a Keplerian ellipse was assumed as unperturbed, reference orbit. Taking the limit $M\rightarrow 0$ implies a breakdown of the validity of such an approximation since  $A_r^{(H\rm N)}\propto v_r\propto \sqrt{M}$; thus, for $M\rightarrow 0$, the Newtonian monopole $A^{(\rm N)} = GM/r^2$ becomes smaller than $A_r^{(H\rm N)}$.

The shift per orbit of the radial velocity \eqi v_{\rho}=\rp{n_{\rm b} a \sI\qua{e\co + \cos\ton{\omega + f}}}{\sqrt{1-e^2}},\eqf which is a typical spectroscopic observable in binaries studies, turns out to be
\begin{align}\Delta v_{\rho}^{(H\rm N)} \lb{vra}\nonumber &= -\rp{2\pi a H_0\sI\co\qua{2e^4 + 2\ton{-1 + \sqrt{1-e^2}} +e^2\ton{-5 + 6\sqrt{1-e^2}} } }{ e^3} =\\ \nonumber \\
& = \rp{5\pi ae H_0\sI\co}{2}\ton{1 + \rp{7}{10}e^2} + \mathcal{O}\ton{e^4}.
\end{align}
Note that \rfr{vra} is independent of $M$: in the limit $M\rightarrow 0$ \rfr{vra} does not vanish. Instead, in the limit $G\rightarrow 0$ \rfr{vra} correctly vanishes because of \rfr{cosmo1}.
\section{Theoretical motivations against the existence of a cosmological $H v_r$ term in the equations of motion of test particles}\lb{vaffa}
Actually, although appealing, the existence of \rfr{accel} does not seem to be justified by a theoretical analysis of particle dynamics rooted in general relativity.

The influence of the cosmic expansion  on the gravitation fields surrounding  individual objects was the subject of several investigations since earlier times \citep{1933MNRAS..93..325M,1945RvMP...17..120E,1946RvMP...18..148E,1954ZPhy..137..595S,2000CQGra..17.2739B,2007PhRvD..75f4011A,2007CQGra..24.5031M,2007PhRvD..75f4031S,2012arXiv1207.3873K,2012MNRAS.422.2931N};
for a recent review covering several aspects, see \citep{2010RvMP...82..169C}. Connections with possible local variations of cosmologically varying constants were elucidated by Shaw and Barrow in \citep{2006PhLB..639..596S,2006PhRvD..73l3505S,2006PhRvD..73l3506S,2007GReGr..39.1235B}.
McVittie \citep{1933MNRAS..93..325M}, working in the framework of general relativity, obtained a new spherically symmetric metric describing a point mass embedded in an expanding spatially-flat universe from a suitable combination of the Schwarzschild and FLRW metrics. After a longstanding debate about its physical interpretation, it seems \citep{2010PhRvD..81j4044K,2011PhRvD..84d4045L} that the McVittie metric actually describes a point mass in an otherwise spatially-flat FLRW universe. However, Kopeikin \citep{2012arXiv1207.3873K}, who quotes Carrera and Giulini \citep{2010RvMP...82..169C}, disagrees with such a conclusion.
Indeed, let us consider the McVittie spacetime metric, written in \virg{physical}, i.e. not comoving, coordinates $\left\{ct,r,\theta,\phi\right\}$ \citep{2010RvMP...82..169C,2010PhRvD..81j4044K,2012MNRAS.422.2931N}
\begin{align}
g_{00} & = 1 - 2\mu -{\mathfrak{h}}^2, \\ \nonumber \\
g_{01} & = \rp{\mathfrak{h}}{\sqrt{1 - 2 \mu}}, \\ \nonumber \\
g_{02} &= g_{03} =0, \\ \nonumber \\
g_{11} & = -\rp{1}{1-2\mu}, \\ \nonumber \\
g_{22} &= -r^2, \\ \nonumber \\
g_{33} & = -r^2\sin^2\theta, \\ \nonumber \\
g_{12} & = g_{13}= g_{23} =0,
\end{align}
where
\begin{align}
\mu &\doteq \rp{GM}{c^2 r}, \\ \nonumber \\
\mathfrak{h} &\doteq \rp{H(t)r }{c}.
\end{align}
The equations of motion can be obtained from the Lagrangian
\eqi\mathcal{L}=-\rp{1}{2}g_{\mu\nu}\dot x^{\mu}\dot x^{\nu},\eqf
where the dot denotes derivation with respect to the proper time $\tau$, as
\eqi \rp{d}{d\tau}\left(\derp{\mathcal{L}}{\dot x^{\mu}}\right) - \derp{\mathcal{L}}{x^{\mu}} = 0.\eqf

As far as the radial equation of motion is concerned, the terms independent of $c$ in $\partial{\mathcal{L}}/\partial r$ are\footnote{As usual, we posed $\theta=\pi/2$.}
\eqi\derp{\mathcal{L}}r = r \dot\phi^2 +\rp{\ton{-GM + H^2 r^3}}{r^2}\dot t^2 - \rp{H\dot r\dot t\ton{1-3\mu}}{\ton{1-2\mu}^{3/2}} +  \mathcal{O}\left(c^{-2}\right).\lb{dLdr}\eqf
The generalized momentum for $r$ is
\eqi p_r\doteq \derp{\mathcal{L}}{\dot r}=\rp{\dot r}{1-2\mu} -\rp{Hr\dot t}{\sqrt{1-2\mu}}.\lb{pr}\eqf

In the weak-field and slow-motion approximation $(1\gg \mu,c\rightarrow\infty,\dot t\rightarrow 1)$, \rfr{dLdr} reduces to
\eqi\derp{\mathcal{L}}r \rightarrow r \dot\phi^2 -\rp{GM}{r^2} + H^2 r - H\dot r,\lb{dLdt2}\eqf while
\rfr{pr} becomes
\eqi\derp{\mathcal{L}}{\dot r}\rightarrow \dot r -Hr.\lb{pr2}\eqf
Thus, \rfr{pr2} yields
\eqi\rp{d}{dt}\ton{\derp{\mathcal{L}}{\dot r}}= \ddot{r} + q H^2 r + H^2 r -H\dot r.\lb{dpr2dt}\eqf
By equating \rfr{dpr2dt} and \rfr{dLdt2} yields
\eqi \ddot{r}- r\dot\phi^2 = -\rp{GM}{r^2} -q H^2 r:\lb{eora?}\eqf
the left-hand-side is nothing but the radial acceleration in polar coordinates, while
both the terms $-H\dot r$ appearing in \rfr{dpr2dt} and \rfr{dLdt2} canceled each other in \rfr{eora?}.
Arakida \citep{2011GReGr..43.2127A} recently studied the effects of the McVittie metric on the gravitational time delay.

Another way to realize that standard general relativity does not predict the existence of \rfr{accel} for a two-body system immersed in a FLRW expanding universe consists of looking at the cosmological  impact on the binary system  through the generalized Jacobi equation as a tidal effect in the local Fermi frame.
\citep{1972GReGr...3..351H,1975ApJ...197..705M,2002CQGra..19.4231C}. The generalized Jacobi equation is suitable for our purposes since it actually takes into account the
relative velocity of the geodesics of the primary and the test particle freely moving in the background FLRW metric \citep{2007CQGra..24.5031M}. It turns out that an acceleration term linear in the velocity of the test particle with respect to the primary is, in general, present in the generalized Jacobi equation: it is proportional to \citep{2002CQGra..19.4231C} ${\mathcal{R}}_{ijk0}V^{j}X^{k},i=1,2,3$, where $X^k$ and $V^j=\dot X^j$ are the Fermi spatial coordinates  of the local Fermi frame, and ${\mathcal{R}}_{\mu\nu\rho\sigma}$ is the projection of the Riemann tensor of the background metric onto the Fermi frame. The problem is that the  components of the Riemann tensor needed to yield \rfr{accel} are vanishing in the FLRW metric \citep{2007CQGra..24.5031M}. It is well known that the standard Jacobi equation, with its ${\mathcal{R}}_{j0k0}, j,k=1,2,3$ components, is able to yield the $H^2$ term of \rfr{accel2} \citep{1998ApJ...503...61C}.

Kopeikin \citep{2012arXiv1207.3873K}, following a cosmological perturbation approach, did not find \rfr{accel} in the equations of motion of a test particle to the Newtonian level. He  \citep{2012arXiv1207.3873K} found \rfr{accel2}, and a  post-Newtonian term  \eqi \bds A^{(H\rm pN)} = H\ton{v/c}^{2}\bds v\lb{accelK}:\eqf its magnitude is quite small in the Solar System.
By employing the methods used in Section \ref{osser}, it is possible to show that \rfr{accelK} causes the following long-term variations of $a$ and $e$
\begin{align}
\dot a^{(H\rm pN)} \lb{rateaK} & = \rp{2H_0 a^3 n^2_{\rm b}\ton{4 - 3\sqrt{1-e^2}}}{c^2\sqrt{1-e^2}}=\rp{2 H_0 a^3 n^2_{\rm b}}{c^2} + \rp{4 H_0 a^3 n^2_{\rm b} e^2}{c^2} + \mathcal{O}\ton{e^4}, \\ \nonumber \\
\dot e^{(H\rm pN)} \lb{rateeK} & = -\rp{4 H_0 a^2 n^2_{\rm b}\sqrt{1-e^2}\ton{-1+\sqrt{1-e^2}}}{c^2 e} = \rp{2 H_0 a^2 n^2_{\rm b} e}{c^2} - \rp{H_0 a^2 n^2_{\rm b} e^3}{2c^2} + \mathcal{O}\ton{e^4}.
\end{align}
Interestingly, \rfr{rateaK}-\rfr{rateeK} do depend on both $G$ and $M$ in such a way that they correctly vanish in the limit $G\rightarrow 0, M\rightarrow 0$.
About vanishing values of $M$, no singularity at all occurs in this case since $A^{(\rm N)}/A^{(H\rm pN)}\propto M^{-1/2}$, so that the Newtonian monopole is always larger than the cosmological acceleration of \rfr{accelK} for $M\rightarrow 0$.
Moreover, \rfr{rateaK} is independent of the semimajor axis of the orbit of the test particle;  to zero order in $e$, it amounts to the product of the Schwarzschild radius $r_s = 2GM/c^2$ of the localized source times the Hubble parameter. For a Sun planet, \rfr{rateaK} yields an increase rate of its semimajor axis as little as $21\ \mu$m cty$^{-1} + \mathcal{O}\ton{e^2}$. According to \rfr{rateeK}, the rate of change of the eccentricity depends on the size of the orbit as $\dot e/e\sim \ton{r_s/a}H$. At first sight, the pericenter precession due to \rfr{accel2}, which amounts to \citep{2007PhRvD..75f4031S}
\eqi\dot\omega^{\ton{H^2\rm N}} = -\rp{3q_0 H_0^2\sqrt{1-e^2}}{2n_{\rm b}},\lb{pre}\eqf gets singular in the limit $G\rightarrow 0$ because of $n_{\rm b}^{-1}$ in \rfr{pre}; actually, it is not so since \citep{2010obco.book.....S}
\eqi -q H^2 \doteq \rp{\ddot{S}}{S} = -\rp{4\pi G}{3}\ton{\rho + 3\rp{p}{c^2}},\eqf where  $p$ is the  pressure of the total cosmic fluid, inclusive of
any dark energy component. In principle, \rfr{pre} has a singularity for $M\rightarrow 0$ as well; actually, it does not make sense to consider such a limit since \rfr{pre} was obtained perturbatively \citep{2007PhRvD..75f4031S} by assuming \rfr{accel2} much smaller than the Newtonian monopole $A^{(\rm N)}$ which, instead, would become smaller than \rfr{accel2} for $M\rightarrow 0$.
The propagation of electromagnetic waves continuously exchanged between two atomic clocks geodesically moving in an expanding FLRW universe retains a cosmological imprint of order $H$ large enough to allow for a possible detection in accurate range-rate Doppler experiments \citep{2012arXiv1207.3873K}. According to Kopeikin \citep{2012arXiv1207.3873K}, it may provide an explanation of the Pioneer anomaly \citep{2010LRR....13....4T} as well.
\section{General considerations on the viability of modified models of gravity from their observational consequences}\lb{conside}
The discussions of the previous sections allow us to provide preliminary general criteria of viability of non-standard models of gravity \citep{2012PhR...513....1C} by inspecting the behaviour of their observable consequences.

Let us suppose to have a given modified model of gravity $\mathfrak{M}$ yielding a test particle extra-acceleration $A^{(\mathfrak{M})}$ that can be viewed as a small correction to $A^{(\rm N)}$ in appropriate circumstances. Depending on the type of model, the parameter(s)  entering $A^{(\mathfrak{M})}$, collectively denoted as $\{\alpha\}$, may or may not contain explicitly  $M$ and $G$. Let us, now, suppose to work out a certain observable effect $X^{(\mathfrak{M})}$, say an orbital extra-precession which slowly alters the otherwise unperturbed Keplerian ellipse. It is clear that, if we ideally switch off gravity,  $X^{(\mathfrak{M})}$ must vanish in the limit $G\rightarrow 0$\textcolor{black}{, at least for a certain class of modified models}. Clearly, the same should occur in the limit $M\rightarrow 0$ as well, unless certain conditions pertaining how $X^{(\mathfrak{M})}$ is calculated must be satisfied. That poses certain basic constraints on the viability of the model $\mathfrak{M}$ and on the nature of its building blocks parameterized by $\{\alpha\}$. It is important to note that it may happen that a well-behaved $A^{(\mathfrak{M})}$ yields a $X^{(\mathfrak{M})}$ which, instead, is not. \textcolor{black}{As far as the condition on vanishing $G$ is concerned, it must be remarked that, strictly speaking, it applies only to a particular class of modified gravity
models where the perturbations  introduced by them to general relativity scale with $G$ itself; it has not necessarily a general validity.}

More precisely, let us consider the case of the Cosmological Constant $\Lambda$ and the orbital effects it causes on a gravitationally bound binary system.  The resulting particle acceleration is \citep{2003CQGra..20.2727K}
\eqi A^{(\Lambda)} = \rp{\Lambda c^2 r}{3}, \lb{accelL}\eqf which yields the orbital precession \citep{2003CQGra..20.2727K,2008AdAst2008E...1I}
\eqi\dot\omega^{(\Lambda)} = \rp{\Lambda c^2\sqrt{1-e^2}}{2 n_{\rm b}}.\lb{precL}\eqf Contrary to what might seem at first sight, \rfr{precL} is, actually, well-behaved with respect to $G$. Indeed, from \citep{2010obco.book.....S}
\eqi\rho_{\Lambda}=\rp{\Lambda c^2}{8\pi G}\eqf and \rfr{critd}, it is possible to write
\eqi\Lambda c^2 =3\Omega_{\Lambda}H^2.\eqf Since, according to \rfr{cosmo1},  $H^2\propto G$ in a spatially flat universe, we see from \rfr{precL} that $\dot\omega^{(\Lambda)}\propto \sqrt{G}$; thus, $\dot\omega^{(\Lambda)}\rightarrow 0$ in the limit $G\rightarrow 0$. The mass $M$ of the central body enters \rfr{precL} as $M^{-1/2}$ through the Keplerian mean motion $n_{\rm b}$. As discussed in the previous Sections, it is meaningless to consider the limit $M\rightarrow 0$ for the precession of \rfr{precL} since it was obtained perturbatively, while $A^{(\rm N)}/A^{(\Lambda)}<1$ for $M\rightarrow 0$.

As another example, let us consider the Dvali-Gabadadze-Porrati (DGP) model \citep{2000PhLB..485..208D}. It predicts an acceleration
\eqi A^{\ton{\rm DGP}} = \mp\ton{\rp{c}{2r_0}}\sqrt{\rp{GM}{r}},\lb{adgp}\eqf
which is well-behaved with respect to the limits $G\rightarrow 0, M\rightarrow 0$; $r_0$ is a free length scale determined by observations of cosmological nature, while the $\mp$ sign depends on the cosmological expansions phases.
As far as the observational consequences of \rfr{adgp} are concerned, it yields an orbital precession \citep{2003PhRvD..67f4002L,2005CQGra..22.5271I}
\eqi \dot\omega^{^{\ton{\rm DGP}}}=\mp\rp{3c}{8r_0} + \mathcal{O}\ton{c^{-2}}\lb{omegadgp}\eqf which is pathologically independent of both $G$ and $M$. The same drawback is common to Galileon-based models as well \citep{2009PhRvD..79f4036N,2010JCAP...08..011B} yielding orbital precession \citep{2012JCAP...07..001I} formally identical to \rfr{omegadgp}.

A further example is given by power-law extra-potentials \citep{2007PhRvL..98m1104A,2009PrPNP..62..102A} \eqi U^{\ton{\mathfrak{K}}} = \rp{\mathfrak{K}}{r^k},\ k>1,\eqf where $\mathfrak{K}$ may or may not depend on $GM$ along with other parameter(s) $\grf{\alpha}$. It can be shown perturbatively that the resulting orbital precessions are \citep{2007PhRvD..75h2001A,Iorio012}
\eqi\dot\omega^{\ton{\mathfrak{K}}}\propto \rp{\mathfrak{K}}{n_{\rm b}};\lb{preck}\eqf if $\mathfrak{K}$ does not contain $G$, an unphysical singularity would appear in \rfr{preck} for $G\rightarrow 0$. From this point of view, the Randall-Sundrum model \citep{1999PhRvL..83.4690R} is well-behaved since $\mathfrak{K}\propto GM\ell^2$ in it, where $\ell$ is the anti-de Sitter (AdS)  curvature scale.

The Kehagias-Sfetsos (KS) \citep{2009PhLB..678..123K} solution of the Ho\v{r}ava-Lifshitz (HL) modified gravity \citep{2009PhRvD..79h4008H,2009JHEP...03..020H} predicts the existence of an additional radial acceleration
\eqi A^{(\rm KS)} \sim \rp{4G^4M^4}{\psi_0 c^6r^5},\lb{accelKS}\eqf where $\psi_0$ is a parameter of the HL model. The orbital precession induced by \rfr{accelKS} is \citep{2010IJMPA..25.5399I}
\eqi\dot\omega^{(\rm KS)}\sim -\rp{3G^{7/2}M^{7/2}}{2\psi_0 c^6 a^{9/2}};\eqf it correctly goes to zero for $G\rightarrow 0, M\rightarrow 0$. The lack of singularity for $M$ is due to the fact that the Newtonian monopole remains always larger than $A^{(\rm KS)}$ for vanishingly small values of $M$.

Clearly, if a model is satisfactory from the point of view illustrated  here, this does not necessarily imply it is really valid  as a good description of (a part of) the physical reality: it has to pass other independent tests. \textcolor{black}{On the other hand, as already remarked, not all modified models induce perturbations scaling with $G$ itself.}
\section{Conclusions}\lb{conclusioni}
We investigated the possibility that the cosmological expansion may impact the orbital motion of a localized two-body system at first order in the Hubble parameter $H$.

 Reasoning classically, intuitive guesses  lead us to postulate, at Newtonian level, the existence of an additional radial acceleration $A^{(H\rm N)}_r$ of order $H$ proportional to the radial velocity $v_r^{(\rm orb)}$ of the proper motion of the test particle with respect to the primary. By considering $A^{(H\rm N)}_r$ as a small correction to the Newtonian monopole $A^{(\rm N)}$, we perturbatively worked out its long-term effects on the  Keplerian orbital elements of the test particle and on its distance from the primary, and, in the case of a binary in the plane of sky, on the projection of its orbit onto the line-of-sight and on its radial velocity. It turned out that both the semimajor axis $a$ and the eccentricity $e$ of the test particle would secularly increase. The analytical expressions of their rates of change are independent of the mass $M$ of the primary, so that they do not vanish in the limit $M\rightarrow 0$, contrary to that one might reasonably expect. Such a feature constitutes a general requisite to be satisfied in the search of potentially viable alternative, i.e. non-cosmological, physical mechanisms able to provide $A_r^{(H\rm N)}$.

 Nonetheless,  the consequences of $A^{(H\rm N)}_r$ would be interesting from a phenomenological point of view since their magnitude is close to the current level of accuracy in determining the planetary orbits in our Solar System.
 %Preliminary analyses of recently published data show a tension among the predicted effects for $A^{(H\rm N)}_r$  and the admissible upper bounds on the %secular rates of change of $a$ and $e$ for Mercury and Mars.

 Then, we looked for theoretical justifications of the existence of $A^{(H\rm N)}_r$ in three different frameworks within general relativity. None of them provided it at Newtonian level, contrary to the known Hooke-like term of order $H^2$. Instead, as recently pointed out in literature, a velocity-dependent acceleration $A^{(H{\rm p N})}$ of order $H$ and directed along the velocity $\bds v$ of the test particle  exists at post-Newtonian level. We perturbatively worked out its orbital effects by finding secular rates of change of $a$ and $e$ proportional to the Schwarschild radius $r_s$ of the primary. They are well-behaved since they correctly vanish in the limit $M\rightarrow 0$. For a planet of our Sun such effects are negligibly small: suffice it to say that the semimajor axis increases at a rate of just 20 $\mu$m per century. We discussed the limits of validity of the orbital precession due to the $H^2$ term. We noticed that the formal singularity occurring in it for $M\rightarrow 0$ actually has no physical meaning since such a limit is beyond the regime of validity of the perturbative calculation yielding the precession itself.

 Finally, we extended some points emerged in dealing with the previous cosmological issues by discussing certain basic criteria of viability that \textcolor{black}{certain classes of} modified models of gravity should generally  meet in view of their predicted observable effects like, e.g., orbital precessions. Given that such modified gravities must reduce to small perturbations of standard Newtonian gravity in appropriate circumstances, their precessions, which slowly alter an otherwise fixed Keplerian ellipse, must necessarily vanish in the limit of no gravity, i.e. for $G\rightarrow 0$\textcolor{black}{, at least as far as modified gravities whose perturbations to general relativity scale with $G$ are concerned; it is not necessarily valid in all cases}. The same should occur also for $M\rightarrow 0$, unless such a limit violates the validity of the pertubative regime in which the precessions are calculated.
 %If a given model provided a precession which is, say, independent of $G$ and/or $M$, caution should be used in further considering that model.
\bibliography{Hubblebib}{}
%-----------------------------------------

\end{document}